\documentclass[10pt,journal]{IEEEtran}
\usepackage{commath}
\usepackage{color}
\usepackage{cite}
\usepackage{amsmath}
\usepackage{latexsym}
\usepackage{graphicx}
\usepackage[FIGTOPCAP,tight,nooneline]{subfigure}

\begin{document}
%
\title{Polarization Engineering in Nano-Scale Waveguides Using Lossless Media}
%
%
%
\author{PoHan~Chang,
        Charles~Lin,
        and~Amr~S~Helmy,~\IEEEmembership{Senior~Member,~IEEE}
\thanks{P. Chang, C. Lin, and A. S. Helmy are with the Department
of Electrical and Computer Engineering, University of Toronto, Toronto,
ON, M5S 3G4 Canada e-mail: (pohan.chang@mail.utoronto.ca).}
\thanks{Manuscript received August 31, 2015}}
%
%

\markboth{}%
{Shell \MakeLowercase{\textit{et al.}}: Bare Demo of IEEEtran.cls for Journals}
%

\maketitle
\begin{abstract}
A device that achieves controllable rotation of the state of polarization by rotating the orientation of the eigenmodes of a waveguide by 45$^{\circ}$ is introduced and analyzed. The device can be implemented using lossless materials on a nanoscale and helps circumvent the inherent polarization dependence of photonic devices realized within the silicon on insulator platform. We propose and evaluate two novel polarization rotator-based schemes to achieve polarization engineering functions: (1) A multi-purpose device, with dimensions on the order of a few wavelengths which can function as a polarization splitter or an arbitrary linear polarization state generator. (2) An energy efficient optical modulator that utilizes eigenmode rotation and epsilon near zero (ENZ) effects to achieve high extinction ratio, polarization insensitive amplitude modulation without the need to sweep the device geometry to match the TE and TM mode attributes. By using indium tin oxide (ITO) as an example for a tunable material, the proposed modulator provides polarization insensitive operation and can be realized with a modulation bandwidth of 112 GHz, a length of 1800 nm an energy per bit of 7.5 $fJ$ and an optical bandwidth of 210 nm.
\end{abstract}

%
\IEEEpeerreviewmaketitle

\section{Introduction}
\IEEEPARstart {I}{n} recent years, silicon-on-insulator (SOI) has been established as the platform of choice for implementing densely integrated photonic devices. For one to be able to truly miniaturize high-performance photonic circuits on SOI, stringent control and management of the state of polarization is required. For instance, the birefringence associated with the waveguides realized in the SOI platform often leads to unwanted polarization-induced dispersion, which can in turn pose limitations on high bit rate systems on chip \cite{Dai13}. Even more detrimental is the polarization dependence of the vast majority of optical devices, which is particularly pronounced on SOI due to the nanoscale confinement of electromagnetic waves. In addition, the manipulation and dynamic control of different polarization states of light is of great significance to numerous applications in optics: the generation of a particular state of polarization and its complement (for example TE and TM linear polarization or left and right circular polarization \cite{Liang14}) is important for enhancing light matter interaction that is essential to many domains of application such as optical sensing, optical signal processing, and the utilization of many quantum mechanical phenomena \cite{Matsuda12,Xu11}.

\par To alleviate the effects of the significant birefringence inherent in SOI waveguides, polarization diversity circuits consisting of a polarization splitter to allow TE and TM signals to be processed separately is often used. This can be utilized in conjunction with polarization rotators, which can ensure that the signal is launched into a given device with the appropriate state of polarization \cite{Barwicz06,Bogaerts07}. The introduction of active materials as part of the rotator-based optical circuit enables electrically-tunable components such as polarization switches, modulators, and dynamic mode converters for quantum as well as classical applications \cite{He14}. Nonetheless, these devices require a relatively large footprint and are therefore not conducive to achieving performance targets, which enable nanoscale photonic device architectures. An alternative route to circumvent birefringence on SOI is to design polarization-insensitive devices, where the properties of the TE and TM modes are matched \cite{Michelotti09,Fujisawa06,Alam12}. This approach often produces complex, narrow-band designs and requires extensive sweeps of the waveguide parameter space. This is an inefficient methodology with no generic design rules and is therefore highly dependent on the specific device considered. More importantly, existing designs for both approaches often incorporate a metal layer to reduce device length, which inevitably leads to partial absorption of the optical signal thereby compromising the device extinction ratio and insertion loss \cite{Michelotti09,Fujisawa06,Alam12}. Therefore, a lossless design platform, where polarization engineering and polarization insensitivity can be readily achieved within the scale of one wavelength without sacrificing the device performance is of great necessity.

\par One prominent example of the trade-offs discussed here is the optical modulator. In order to transform electrical signals into optical ones on a nanoscale, most of the modulators currently reported are based on plasmonic or slot waveguide architectures, thus operate only for TM modes. Even though some carefully optimized structures have been proposed to equalize the modulation loss of both polarizations \cite{Liu12,hu14,Zhu:14}, a general and simple design methodology is still lacking, where polarization insensitive devices can be readily achieved without necessarily sacrificing the device performance.

\par In this work, we demonstrate and evaluate a versatile, efficient and nanoscale polarization rotator which can serve as the basis of an optimum technology for polarization management. The examples provided here will be implemented in the SOI platform, but are equally applicable to other platforms. The design methodology is based on manipulating the coupling and superposition of the two eigenmodes of a waveguide, which are oriented at 45$^{\circ}$ with respect to the wafer axis. First, we present how the introduction of material asymmetry can reduce the footprint of a metal-free polarization rotator without sacrificing its polarization or conversion efficiency. Next, two demonstrations of the rotator-based optical device designs are introduced and evaluated in this work: (1) By utilizing a directional coupler system consisting of Si nanowire and the rotator described here, we show that the interference of the generated supermodes could tailor the output state of polarization to function as either a polarization splitter or a dual-polarization generator. Such capability also offers great flexibility to serve as a polarization (de)multiplexer. (2) By incorporating active materials that support epsilon-near-zero effects such as indium tin oxide (ITO), the same rotator could be utilized as a polarization-insensitive modulator. Due to the equal decomposition of the input state of polarization into the two rotated eigenmode modes (both rotated by 45$^{\circ}$), judicial design is then readily available and waveguide geometry sweeping is not required to find the crossing of TE and TM attenuations. The ENZ effect plays a dual role in this modulator design: eigenmode rotation as well as high extinction amplitude switching \cite{Lu12}. The significance and novelty of this architecture is that the aforementioned devices can all be realized using wavelength-scale cross section (100s of nm) and a length of a few wavelengths (1000s of nm). The notion of polarization engineering in nanoscale, lossless waveguide configurations, where various polarization combinations can be dynamically produced, rotated, and eliminated, has the potential to usher in a new generation of truly nanoscale photonic circuits on SOI.


\section{L-slot Polarization Rotator}\label{PR}
\subsection{Polarization rotation in L-slot waveguides}
\begin{figure}[t!]
\centering
\includegraphics[width=0.33\textwidth]{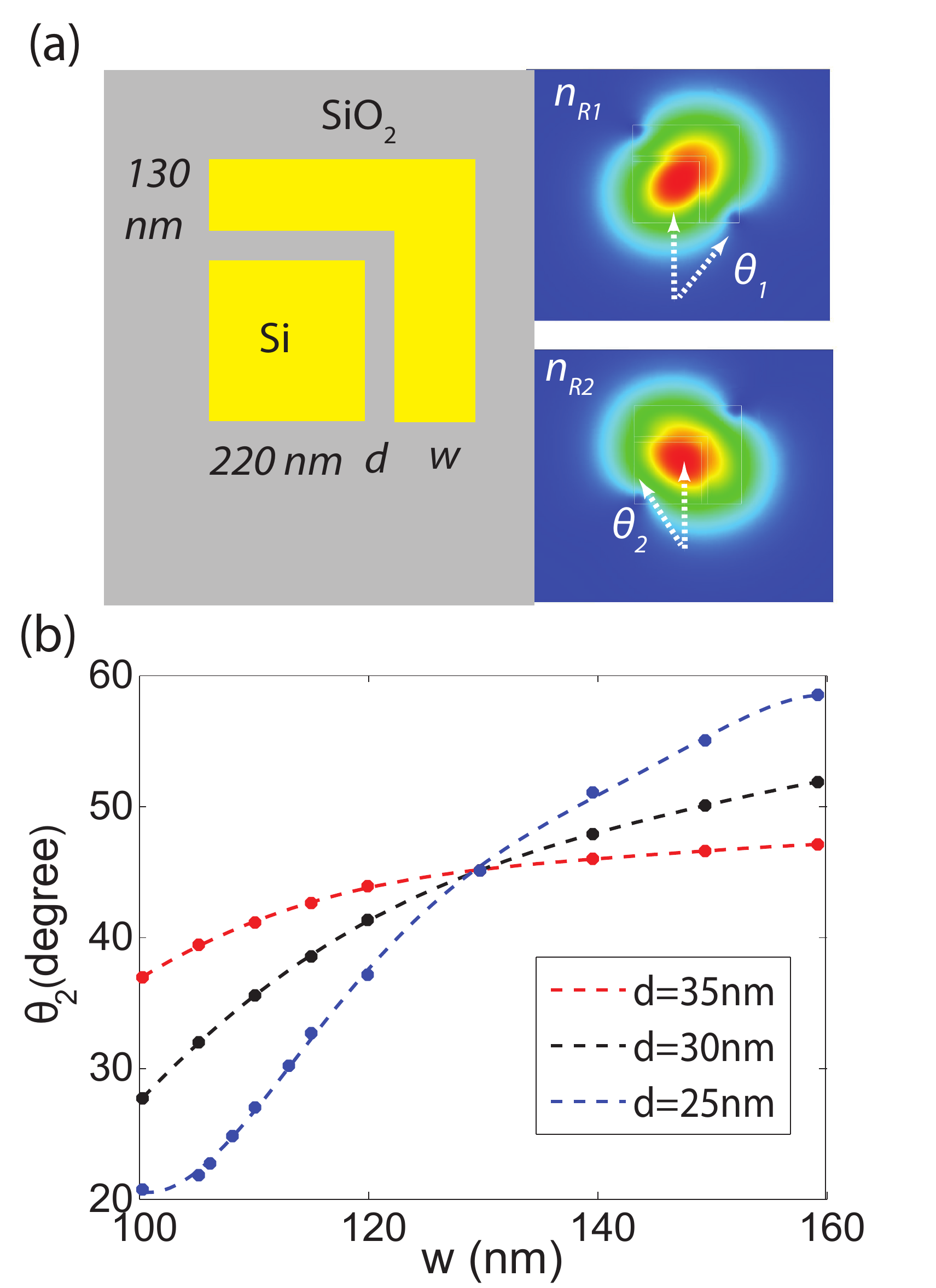}
\caption{\textbf{L-slot polarization rotator.} (a) Schematic of the rotator waveguide and the associated magnetic field profiles of the two eigenmodes, where $n_{R1}$ and $n_{R2}$ denote the effective indices of the two rotated eigenmodes. (b) Angle of orientation ($\theta_2$) as function of the width of the L-shaped cladding ($\theta_1$+$\theta_2$=90$^{\circ}$). Thicker low index region leads to more relaxed geometry intolerance. The width and height of the core waveguide are 220 nm.}\label{She}
\end{figure}

The proposed rotator architecture is based on a L-shaped slot waveguide as shown in Fig. \ref{She}(a). The core nanowire and the outer L-shaped cladding are made in this example out of silicon (yellow). However any high index material will function just as well. In the L-slot region a low index material (in this case it will be taken as $SiO_{2}$) is sandwiched between the two silicon layers to support slot modes. The structure is then covered with silica to reduce the index contrast thereby increasing the degree of eigenmode rotation \cite{Wang08}. The implementation of our rotator can be realized through the fabrication processes of a ridge Si waveguide, followed by a conformal deposition of a low index dielectric and a high index dielectric such as amorphous Si \cite{Morichetti07}. After another lithography step, these two newly deposited layers can be selectively etched to provide asymmetry ridge analyzed in this work. Due to the inherent symmetry breaking that is intrinsic to this waveguide, the predominant transverse polarization component is oriented at an angle $\theta$ with respect to the original optical axis, where the $x$ and $y$ axis are parallel and orthogonal to the sample surface respectively as can be seen in the figure \cite{Wang08}:
\begin{equation}
\tan\theta=\iint{\frac{\varepsilon\left(x,y\right) H_{x}^{2}\left(x,y\right)}{\varepsilon\left(x,y\right) H_{y}^2\left(x,y\right)}dxdy},\label{the}
\end{equation}
where $\varepsilon$ is the real part of permittivity of the materials used. The rotation angle in (\ref{the}) can also be visualized as the angle of orientation of the magnetic fields of the eigenmodes with respect to y-axis as shown in Fig. \ref{She}(a). For instance, the rotated eigenmodes become TE and TM polarized modes as $\theta$=0 ($H_{x}$=0) and $\theta$=90$^{\circ}$ ($H_{y}$=0), respectively. The angle of orientations of the eigenmodes shown in Fig. \ref{She}(b) are simulated using Comsol 3.5a RF module, where the domain and the boundary conditions of the simulation are set to be 5 $\mu$m$\times$5 $\mu$m using the scattering boundary condition. The refractive indices of silicon and silica are taken as 3.48 and 1.442 respectively at wavelength of 1550nm. 
\par In cases where mode interference is the basis of a polarization rotator, the eigenmodes rotation angles serve as an essential attribute that determines the performance. This is because any input state of polarization is decomposed into the two rotated eigenmodes upon propagation. Maximal polarization conversion is achieved when both eigenmodes undergo 45$^{\circ}$ rotation. This is generally accomplished by highly asymmetric waveguide geometries \cite{Gao13,Kim15} or by utilizing a diagonal symmetry of the waveguide structure \cite{Fukuda08}. However, asymmetric waveguide rotators require metal layers in order to break the symmetry of the original TE and TM modes in a dramatic fashion. This approach suffers from sever fabrication intolerance and plasmonic-induced insertion losses. Diagonally symmetric waveguide rotators, on the other hand, represent a more advantageous platform because of its versatility, lower losses, and compatibility with various waveguide architectures \cite{Fukuda08,Jin14}. In this work, where we use L-slot rotators, the diagonal symmetry condition and the corresponding 45$^{\circ}$ orientation angle can be intuitively engineered by equating the height and width of each waveguide layer with the values that achieve resonance in the vertical and horizontal slot modes. The device length can then be calculated by $L_{\pi}$= $(\lambda$/2)($n_{R2}$-$n_{R1}$). For instance, if one utilizes waveguide dimensions for the core, slot, and cladding of 220, 30, and 110 nm respectively, the rotator coupling length is calculated to be 24 $\mu$m.
\par At wavelengths of operation in the vicinity of 1550 nm, the diagonal slot waveguide mechanism is particularly suitable for implementation on SOI, where the Si thickness is 220 nm. This is because the fundamental total internal reflection (TIR) mode would be cut off in the core nanowire for the dimensions we utilize. This cutoff condition ensures that most of the energy of any input polarization would couple to the two rotated eigenmodes. In addition, by increasing the dimensions of the low index material in the slot and the L-shaped top cladding the rotation angle can approach 45$^{\circ}$ even without diagonal symmetry, which in turn drastically improves the tolerance to fabrication errors. If one allows for a design specification such that the required state of polarization rotation in the waveguide should be 45$\pm$5$^{\circ}$ for device operation to fall within the allowed tolerance level, it can be seen in Fig. \ref{She} (b) that in case the slot width $d$ is increased from 25 to 35 nm, the range of widths for the top layer $w$ which maintains operation within specifications can exceed 40 nm. This level of tolerance is well within many existing fabrication processes.
\begin{figure}[t!]
\centering
\includegraphics[width=0.49\textwidth]{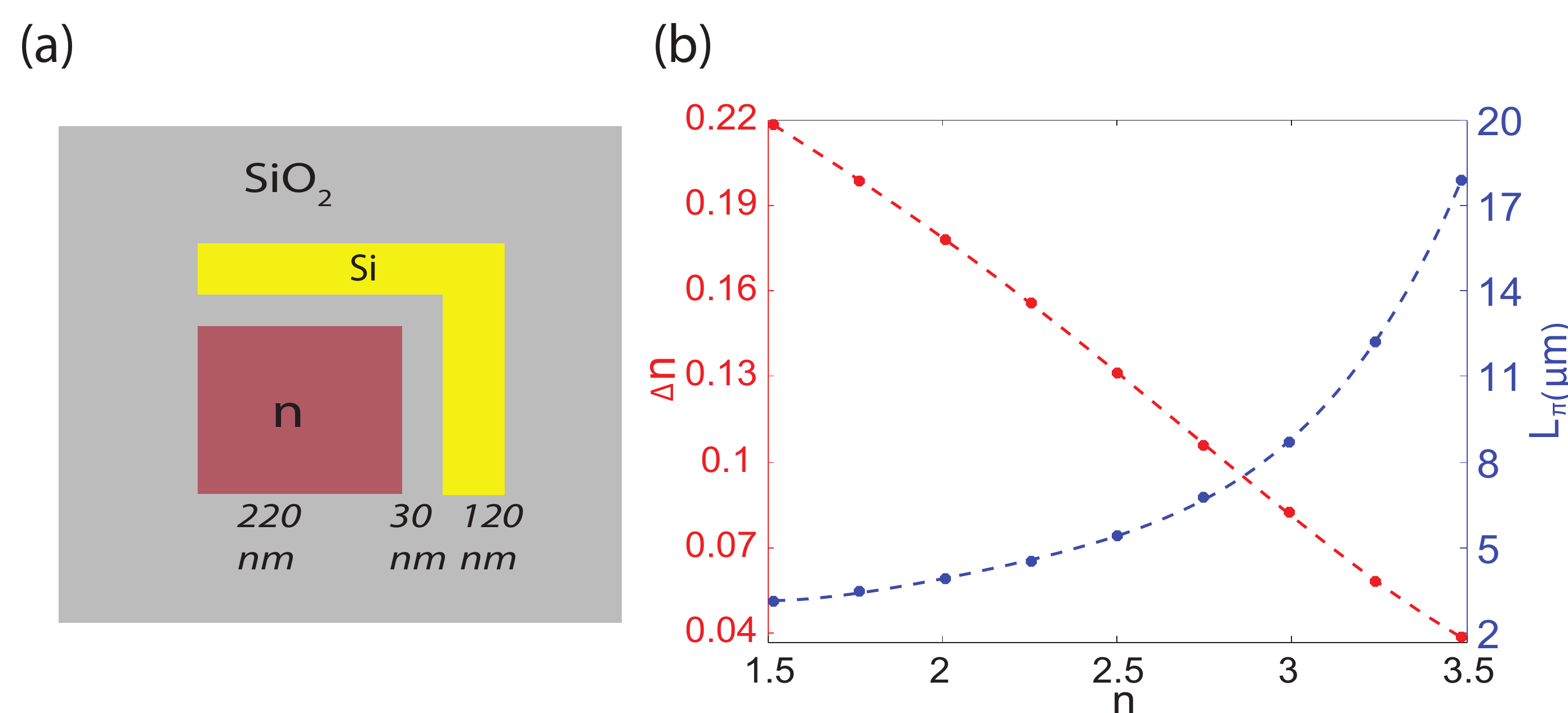}
\caption{\textbf{Toward nanoscale rotator.} (a) Schematic of the designed polarization rotator. The nanowire core is replaced by other high index material. (b) Birefringence ($\Delta$n=$n_{R2}$-$n_{R1}$) and the corresponding coupling length ($L_{\pi}$) as functions of the refractive index of the core material. }
\label{core}
\end{figure}
\subsection{Sub-micron polarization rotators in lossless media}
The proposed L-slot rotator can also be realized within a micrometer footprint in order to better meet the miniaturization demands of optical components. This demand is driven by the need to enhance the integration density of photonic circuits. Enhancing the index difference between the core and cladding materials as can be seen in Fig. \ref{core}(a) increases the asymmetry of the waveguide structure along the diagonal direction thereby increasing the birefringence among the two rotated eigenmodes. As shown in Fig. \ref{core}(b), the coupling length can be achieved on a micrometer scale ($<$5 $\mu$m) as the core index is reduced below 2.5. This can also be achieved if the silicon core is replaced by $Si_{3}N_{4}$ or $HfO_{2}$. Such a feature provides the L-slot waveguide structure with tremendous design flexibility and allows it to be  implemented within a diverse set of material platforms (such as III-V semiconductors, SOI or $Si_{3}N_{4}$). Even though the transition from a slot to TIR mode would give rise to larger mode area of the two rotated modes, the overall device footprint remains on the nanoscale in the transverse direction with device lengths on the order of 1-3 $\mu$m.
\begin{figure*}[t!]
\centering
\includegraphics[width=0.8\textwidth]{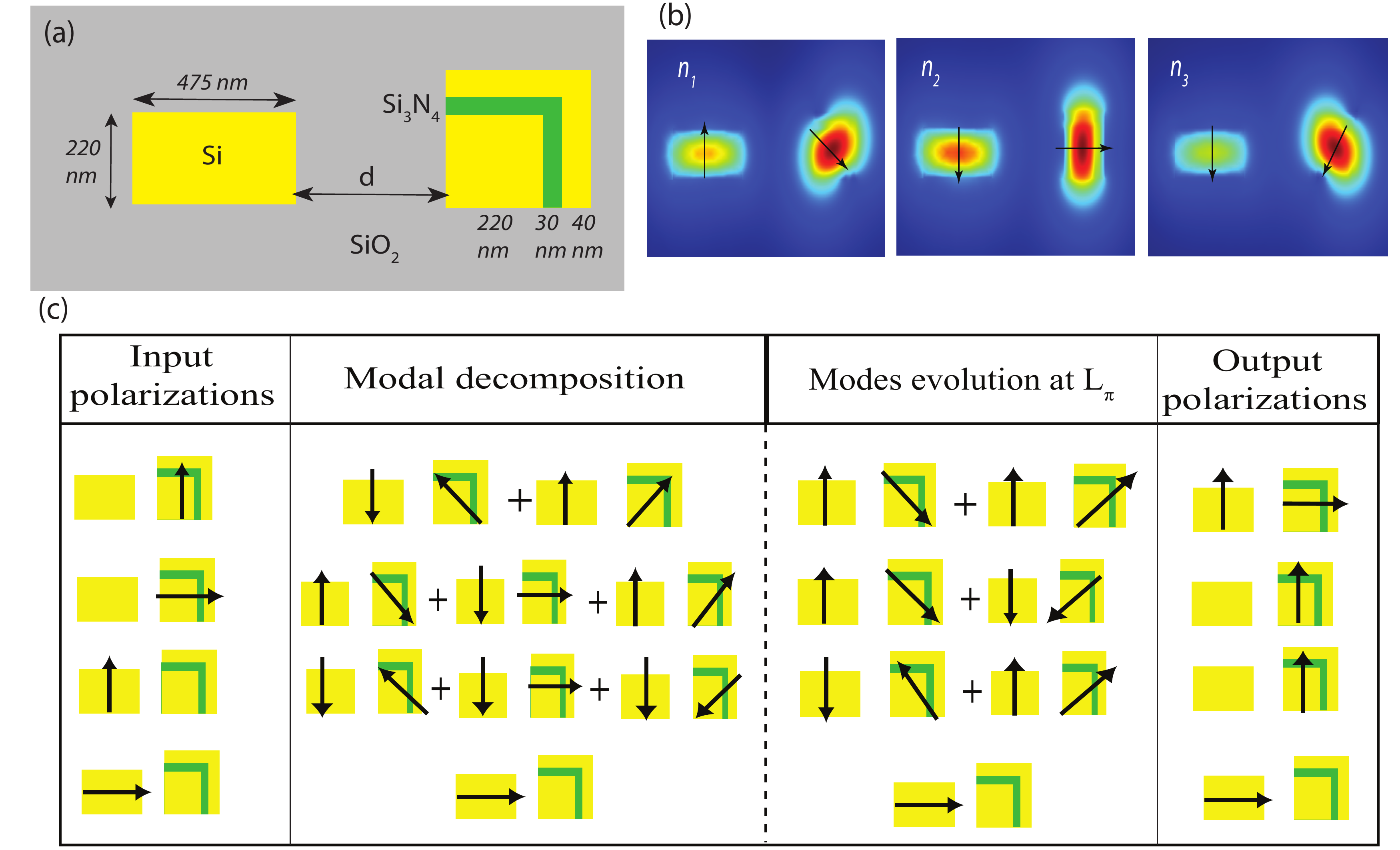}
\caption{\textbf{Rotator coupler and summary of polarization engineering.} (a) Schematic of the proposed rotator coupler consisting of a silicon nanowire ($n_{TM}$=1.752, $n_{TE}$=2.6) and a rotator waveguide ($n_{R1}$=1.736, and $n_{R2}$=1.77). (b) Magnetic field profiles of the supermodes under resonant condition (\ref{resonance}), where the arrows represent the corresponding phase of the electric fields. Assuming separation distance $d$=550 nm, the effective indices of the three supermodes are: $n_{1}$=1.729, $n_{2}$=1.753, and $n_{3}$=1.776, respectively. (c) Summary of the polarization evolution under different input polarizations.}\label{evolution}
\end{figure*}
\par The underlying rotation and miniaturization mechanism is based on designing waveguide with diagonal symmetry using lossless media. These attributes together with the performance of this approach, as will be demonstrated, distinguishes it from plasmonic based rotators as those suffer from significant insertion loss \cite{Gao13,Jin14}. The mode interference nature is also not limited by scattering losses. Rotators that are based on mode conversion require a significantly larger footprint to adiabatically transform modes using gratings or other structures, with sizeable scattering losses associated \cite{Zhang11,Caspers12}. Furthermore, the reduction of the device coupling length, from slot to TIR mode by decreasing the index of the core as seen in Fig. \ref{core}(b), does not sacrifice the polarization conversion efficiency in any way since both eigenmodes simultaneously achieve 45$^\circ$ rotation. This in turn eliminates the usual tradeoff between polarization conversion efficiency and device length. For instance, highly asymmetric dielectric-based designs have been proposed to effectively shrink the rotator size \cite{Velasco12}. However, the lack of diagonal symmetry renders the two rotated eignemodes in these proposals, hardly orthogonal to each other thereby reducing the conversion efficiency as the modes undergo interference.
\section{Rotator coupler}
The polarization management device proposed here comprises of a L-slot rotator waveguide, which is included in an asymmetric directional coupler. In this coupler, the rotator waveguide is separated by a distance $d$ from a silicon nanowire as can be seen in Fig. \ref{evolution}(a). Unlike the normal phase matching condition present for most of the coupled waveguide systems, in this design we tailor the effective index of one of the eigenmode in the silicon nanowire, the TM mode for instance, to match the average indices of the two rotated eigenmodes of the rotator waveguide in Fig. \ref{She}(a):
\begin{equation}
n_{TM}=\frac{(n_{R1}+n_{R2})}{2}.\label{resonance}
\end{equation}

\par Figure \ref{evolution}(b) plots the supermodes for the condition in (\ref{resonance}). It is interesting to note that in addition to the rotated supermodes $n_{1}$ and $n_{3}$, there exists a TE-TM supermode $n_{2}$ whose effective index is also the average of the other rotated modes: $n_{2}$$\simeq$($n_{1}$+$n_{3}$)/2. The state of polarization of the output of such a coupler could thus be tailored by the interference among these supermodes. It is important to note that (\ref{resonance}) is analogous to the resonance condition required to maximize power transfer in three-core couplers \cite{Donnelly86,Yue09,Lou12}. The only difference is that the even and odd supermodes excited in the three-core couplers have the same orientation of polarization (TE or TM), while in our rotator coupler the direction of polarization states in the rotator waveguide of each supermode are not the same.

\par We now consider the case when the incoming light is launched into the rotator waveguide. An incident TM mode could excite mode-1 and mode-3 and as the modes travel a distance $L_{\pi}$ ($\lambda$/2($n_{3}$-$n_{1}$)), they would be out of phase with each other thereby generating a TM polarized mode in the silicon nanowire and a TE polarized mode in the rotator waveguide. In contrast to the previously proposed polarization rotator splitter (PRS) \cite{Dai11,Liu11,Wang14}, which rotates one polarization and maintains the other polarization, the coupler here simultaneously generates opposite polarizations with equal power splitting. This is particularly useful in some quantum optical applications, which require the generation of dual polarizations from a source with a single polarization. For light incidence with TE polarization, on the other hand, the three supermodes of the waveguide system would all be excited and produce a TM polarized mode in the rotator waveguide over a distance $L_{\pi}$. In addition, after a distance of 2$L_{\pi}$ it can also produce a TM polarized mode in the silicon nanowire and TE polarized mode in the rotator waveguide.
\par For the cases, where the light is launched into the silicon nanowire side, the TE polarized light would pass through the silicon nanowire with no coupling to the rotator waveguide. This is because the resonance condition is matched for TM only. For the TM input polarized light, in contrast, all the three supermodes of the structure would be excited and the same polarization would be imaged at the rotator waveguide over a distance $L_{\pi}$. Under these operating conditions the proposed coupler could separate different polarizations thus functioning as a polarization beam splitter (PBS). The overall mode evolution and polarization evolution of the modes are summarized in Fig. \ref{evolution}(c).

\par It is important to highlight that the rotator-based coupler is also beneficial for device miniaturization. The refractive index difference between $n_{1}$ and $n_{3}$ in Fig. \ref{evolution}(b) is larger than the difference between $n_{R1}$ and $n_{R2}$ in a stand alone rotator, seen in Fig. \ref{She}(a). In the example depicted in Fig. \ref{evolution}(a), the coupling length for the rotator coupler (16.4 $\mu$m) is 30$\%$ shorter than the rotator without coupling (22.8 $\mu$m). However in order to generate both polarizations utilizing a single polarization source, a cascade of power splitters followed by a polarization rotator is usually required. As such the overall device length is usually much longer than the 22.8 $\mu$m length. By utilizing the designs described here the device length could be greatly reduced by merging two components into one, while effectively reducing the device length of the rotator at the same time.
\par
\begin{figure}[t!]
\centering
\includegraphics[width=0.5\textwidth]{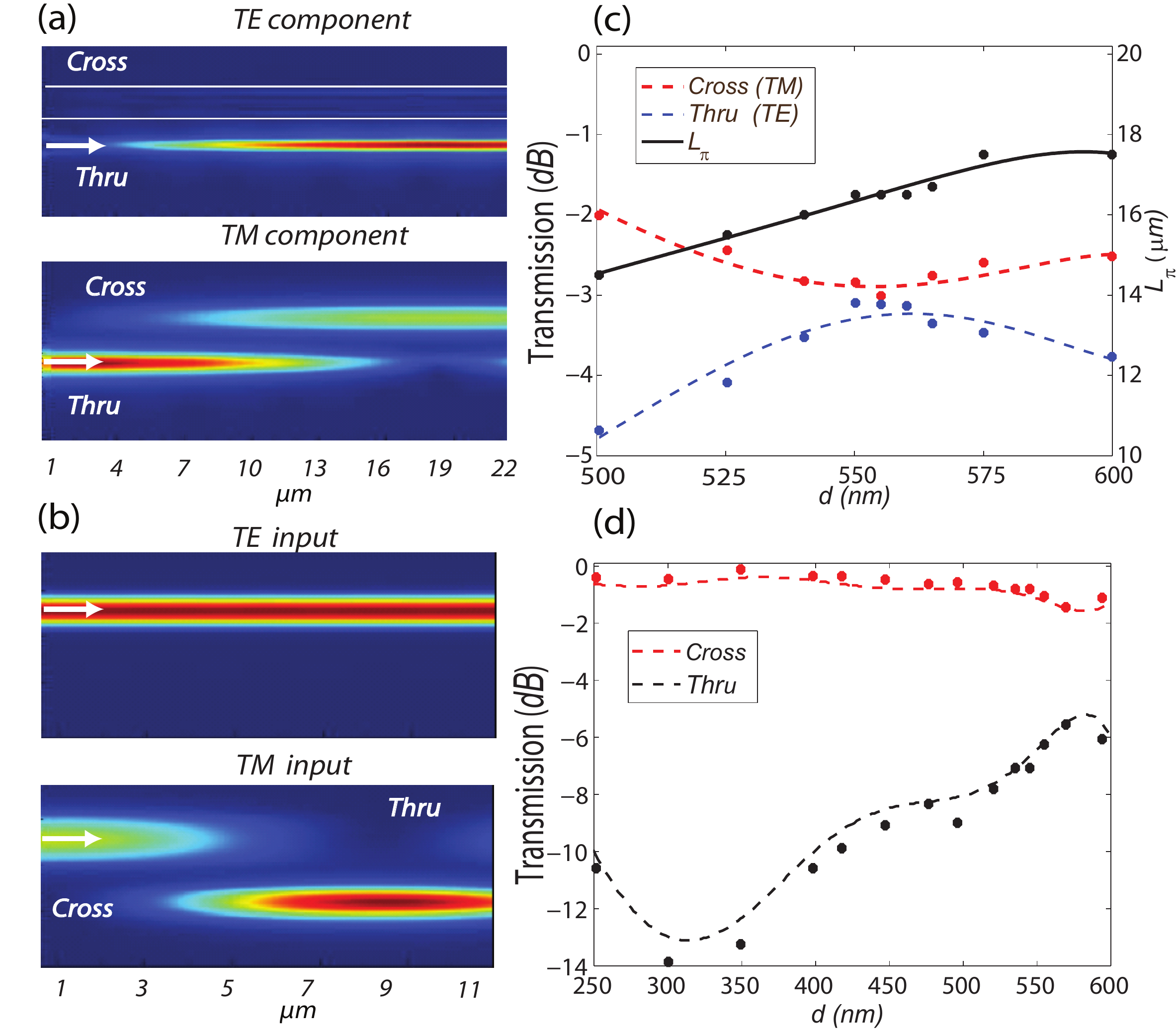}
\caption{\textbf{Mode conversion in the designed rotator coupler.} (a) TM incidence on the rotator waveguide (row 1 of Fig. \ref{evolution}(c)). The top and the bottom are TE and TM components of the electric field intensities. (b) Light incidence on the silicon nanowire. Top is under TE incidence (row 4 of Fig. \ref{evolution}(c)) and the bottom (row 3 of Fig. \ref{evolution}(c)) is under TM incidence. The white arrows indicate the direction propagations of light. (c) and (d) are the transmittances associated with (a) and (b) at different waveguide separations. }\label{simulation}
\end{figure}

 The mode evolution, which is simulated using eigenmode expansion method (EME) \cite{lumerical}, is shown in Figure \ref{simulation}, where the domain and the boundary condition are set to be 4.2$\mu$m$\times$1.7$\mu$m and metal. The plot in Fig \ref{simulation}(a) shows the TE (top) and TM (bottom) field components when a TM-polarized beam is launched into the rotator waveguide (Thru port) at $d=$560 nm. In this case at a distance $L_{\pi}$ the TE and TM polarizations exist the device at the Thru and Cross ports, respectively. The power transmitted at the output ports is also plotted in Fig. \ref{simulation}(c). This result is obtained through an S-matrix calculation of the overlap of the modes with the incident beam. As the waveguide separation varies from 500 nm to 600 nm it is found that a 3 dB polarization splitting can be achieved at $d=$560 nm. This value is consistent with the condition given in (\ref{resonance}). For a TE polarized beam launched into the silicon nanowire, as shown in Fig. \ref{simulation}(b), the TE polarization propagates through the waveguide with negligible coupling to the Cross port. When a TM polarized light is launched into the silicon nanowire, however, it mainly couples to the Cross port at the output. As such the dimensions of this device serves to get it to operate as an effective polarization states separator.

\par For polarization splitting in which TM is launched into the silicon nanowire as can be seen in Fig. \ref{simulation}(d), there is a portion of the input energy that remains in the Thru port and the Cross port is composed of mixed polarization state. The deviation from ideal splitting is due to the imperfect coupling coefficients between the input beams and the supermodes of the structure. For instance, at $d=$560 nm EME predicts that the TM mode excitation via the rotator waveguide gives rise to 52$\%$ power transfer to mode-1 and 45$\%$ to mode-3. In the case of TM excitation via the silicon waveguide EME predicts 32$\%$ power transfer to mode-1 and 19$\%$ to mode-3. The unequal level of excitation of mode-1 and mode-3 account for the low extinction ratio of the polarization states at the output. Nevertheless, it is interesting to note that the extinction ratio could be enhanced by adjusting the separation between the two elements comprising the rotator coupler. The plot in Fig. \ref{simulation}(d) shows that a maximum polarization extinction ratio at the output can be reached at $d=$300 nm, where the coupling length is further minimized to 7 $ \mu$m as can be seen in Fig. \ref{simulation}(b). Therefore adjusting the separation between the two waveguides can be used to alleviate the limitation associated with unequal power coupling among different supermodes by tuning the effective indices for a desired interference pattern. In contrast to the case studied above for $d=$560 nm, the case of $d=$300 nm, where $n_{1}$, $n_{2}$, $n_{3}$ are 1.807, 1.761, and 1.7 respectively it is found that $n_{2}$$\simeq$($n_{1}$+$n_{3}$)/2 is no longer satisfied.

\section{Rotator-based Modulator}
\subsection{L-Slot modulator}
\par
\begin{figure}[b!]
\centering
\includegraphics[width=0.45\textwidth]{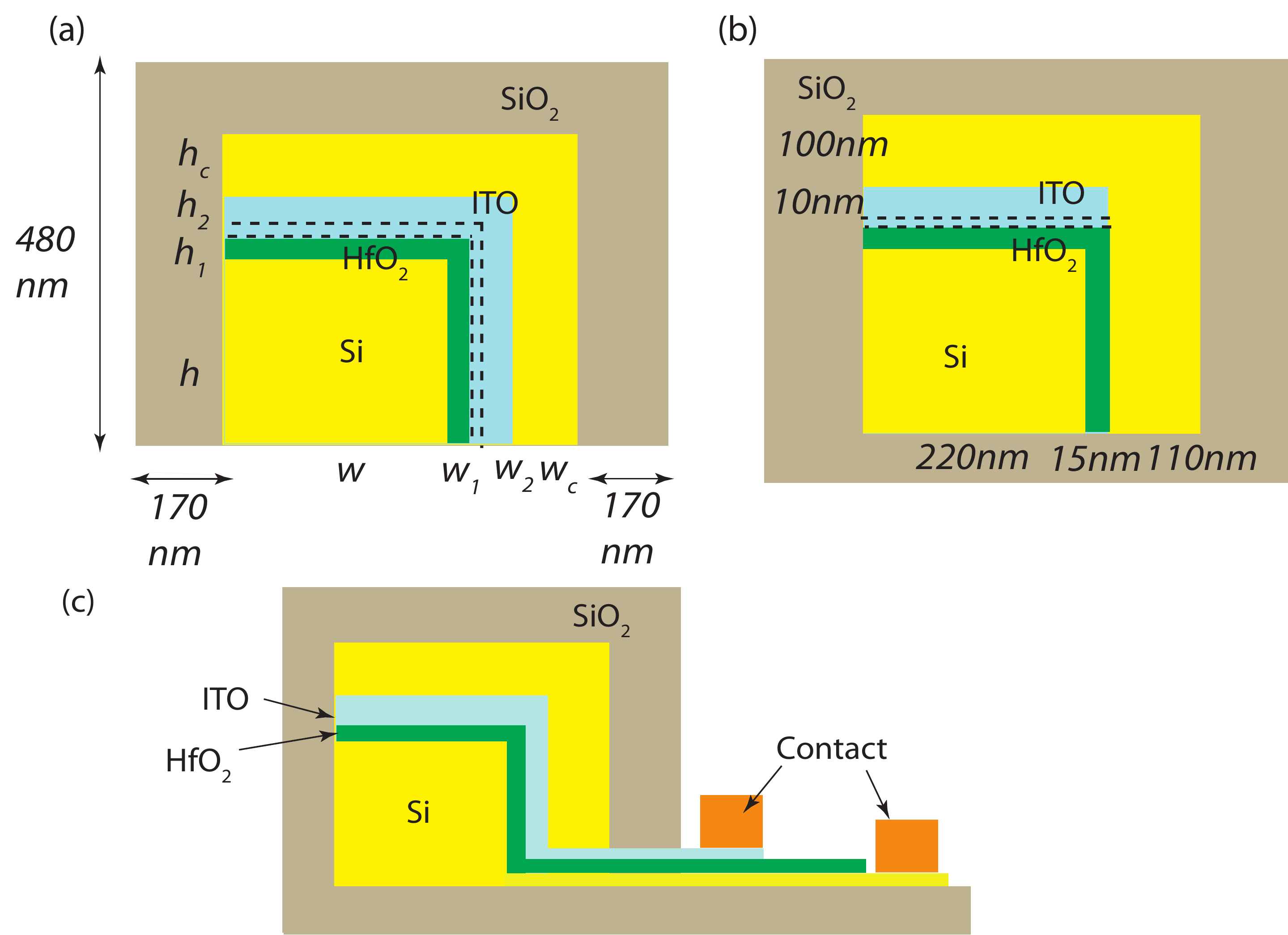}
\caption{\textbf{ITO-based modulator.} (a) Cross section illustration of the L-slot ITO modulator. The dimension parameters are of the followings: $h=w=$220 nm, $h_{1,2}=w_{1,2}=$10 nm, and $h_{c}=w_{c}=$110 nm. The whole device is of 690 nm width and 480 nm height. The asymmetry of the outer silica is to pronounce the ENZ effect on 45$^{\circ}$ rotation. (b) Cross section of the modulator geometry operated under non-ENZ absorption ($\mu$=-5V). The effective accumulation layer thickness is 1 nm at ITO ($n$=2) and HfO2 ($n$=2) boundary, as indicated by the dashed region. (c) Schematic biasing configuration of the modulator. \label{schematic}}%
\end{figure}

 In this section we demonstrate how the addition of a tunable material (ITO for example) could be integrated into the rotator waveguide described above to achieve the function of polarization insensitive modulator. Based on the previous L-slot architecture, ITO is incorporated into the low index region where it has appreciable overlap with the slot modes as schematically shown in Fig. \ref{schematic}. The core and the outer L-shaped cladding are both made of silicon (yellow) and in the L-slot region ITO (blue) is placed on top of high index $HfO_{2}$ (green) to help support high bias voltage before breakdown. In the near infrared wavelength regime, where optical communications systems operate, ITO is emerging as a strong tunable material because it features large index modulation and epsilon near zero (ENZ) effects \cite{Rhodes06,Atwater10,Naik13}. Unity index change allows for dramatic switching between the guided modes thereby enabling larger modulation depths ($>$3 $ dB/\mu$m) with a nanoscale footprint \cite{Melikyan11,Zayats12,Sorger12,Atwater14,Kim15}. On the other hand, the utilization of this L-slot architecture the enhancement of the electric field due to ENZ effects enables strong waveguide absorption \cite{Lu12,Vasudev13,Zhao15}, where the strong intensity of the electric field can be well overlapped with the absorbing materials. Recently, by engineering field distribution with plasmonic modes, ENZ-assisted modulator was reported to enable extremely high figures of merit (FOM) while effectively alleviating the tradeoffs associated with modulators performance \cite{Charles15}. As such it is evident that ITO offers attractive and versatile material properties for both electro-optic and electro-absorption modulators.
By injecting carriers into the modulator structure described here, in this study we take the initial doping concentration of ITO to be $N=10^{19}$ cm$^{-3}$ and the effective accumulation layer thickness to be 1 nm at the interface of ITO and $HfO_{2}$  as can be seen in Fig. \ref{schematic}(a) \cite{Vasudev13}. The refractive index of this interfacial layer can be described by the Drude model, where the ENZ effect ($n=0.55-0.52i$) takes place under a negative bias voltage of $\mu=$-2.3 V ($\Delta n=6.3\times10^{20}cm^{-3})$ at wavelength of 1550 nm \cite{Michelotti09,Vasudev13}. It should be noted that even though -2.3 V is assumed in this work,  this value may vary based on the fabrications and the characterizations of the electrodes of the modulator.

\begin{figure}[t!]
\centering
\includegraphics[width=0.49\textwidth]{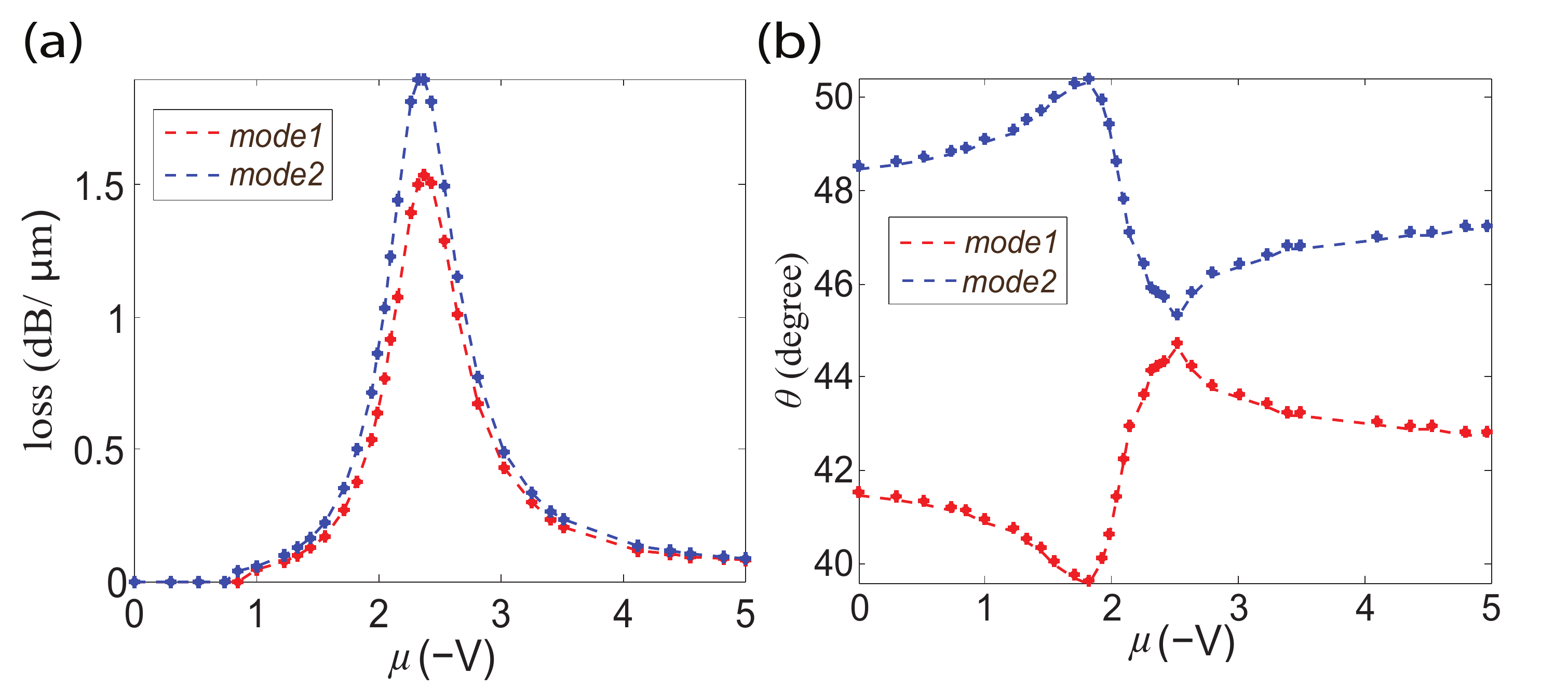}
\caption{\textbf{Mode characteristics of the rotator ITO modulator.} \textbf{(a)} Propagation losses of the two rotated eigenmodes as function of gate voltage at $\lambda=$ 1550 nm, where the losses reach maximum at $\mu=$-2.3 V (ENZ). \textbf{(b)} Angle of orientations of the eigenmodes. $\theta_{1}+\theta_{2}=90^{\circ}$ holds under all voltage values and particularly exhibits 45$^{\circ}$ orientation at $\mu=$-2.3 V. \label{loss}}%
\end{figure}

Since the eigenmodes in the modulator waveguide are rotated with respect to the device axis, the input TE and TM polarizations decompose into a combination of the two rotated eigenmodes:
\begin{equation}
\left[\begin{matrix}
{{H}_{x}}(TM)\\
{{H}_{y}}(TE) \\
\end{matrix} \right]=\left[\begin{matrix}
\cos\theta &\sin\theta \\
-\sin\theta&\cos\theta \\
\end{matrix}\right]\left[\begin{matrix}
{{H}_{1}}\\
{{H}_{2}}
\end{matrix}\right],
\end{equation}
where $\theta$ is given by (\ref{the}).

The intensity of the input polarizations inside the modulator, which are proportional to the square of the magnetic field ($|H^2|$) for a nonmagnetic material, could be expressed as:
\begin{equation}
\left[\begin{matrix}
{{P}_{TM}}\\
{{P}_{TE}} \\
\end{matrix} \right]=\left[\begin{matrix}
\cos^2\theta &\sin^2\theta \\
\sin^2\theta&\cos^2\theta \\
\end{matrix}\right]\left[\begin{matrix}
{{P}_{1}}\\
{{P}_{2}}
\end{matrix}\right],\label{orthogonal}
\end{equation}
where the cross terms vanish due to the orthogonality of the eigenmodes, as will be further discussed. Based on (\ref{orthogonal}), polarization insensitive modulators can be implemented by designing the waveguide to have the eigenmodes oriented at an angle $\theta=$45$^{\circ}$. This is in contrast to the conventional approach of designing the waveguide of a polarization insensitive modulator, which ensures that the loss of $P_1$ is equal to that of $P_2$. The latter generally requires time-consuming geometry sweeping. An optimum design methodology would be one, where an input field with an arbitrary state of linear polarization is split equally into the two eignemodes as $P_{TE,TM}=\frac{1}{2}P_{1}+\frac{1}{2}P_{2}$. Both TE and TM would then share the same attenuation, which is dominated mostly by the eigenmode with the largest loss. Next, we shall provide an illustrative example employing this optimum design methodology, while utilizing a structure, where the eigenmode are oriented at an angle $\theta=$45$^{\circ}$.

\par The design utilizes the L-slot rotator proposed in section \ref{PR} by incorporating ITO as the tunable material. For the waveguide dimensions shown in Fig. \ref{schematic}(a), the plots in Fig \ref{loss} show the loss and angle of orientation of the eigenmodes at different bias voltages ($\mu$). It is seen that the two eigenmodes exhibit peak absorption losses of 1.9 and 1.5 $dB/\mu$m and 45$^{\circ}$ orientation angle at a bias of $\mu=$-2.3 V, which is within the ENZ operating region. This absorption behavior and eigenmode orientation are advantageous as they enable such a waveguide to serve as polarization insensitive absorption modulator. In contrast, the losses decrease dramatically to 0.057 $dB/\mu$m and 0.046 $dB/\mu$m at $\mu=$-1 V. The swing between these two dramatic absorption regimes provides large modulation depths by tuning the bias level of the ITO layer.

\begin{figure}[b!]
\centering
\includegraphics[width=0.32\textwidth]{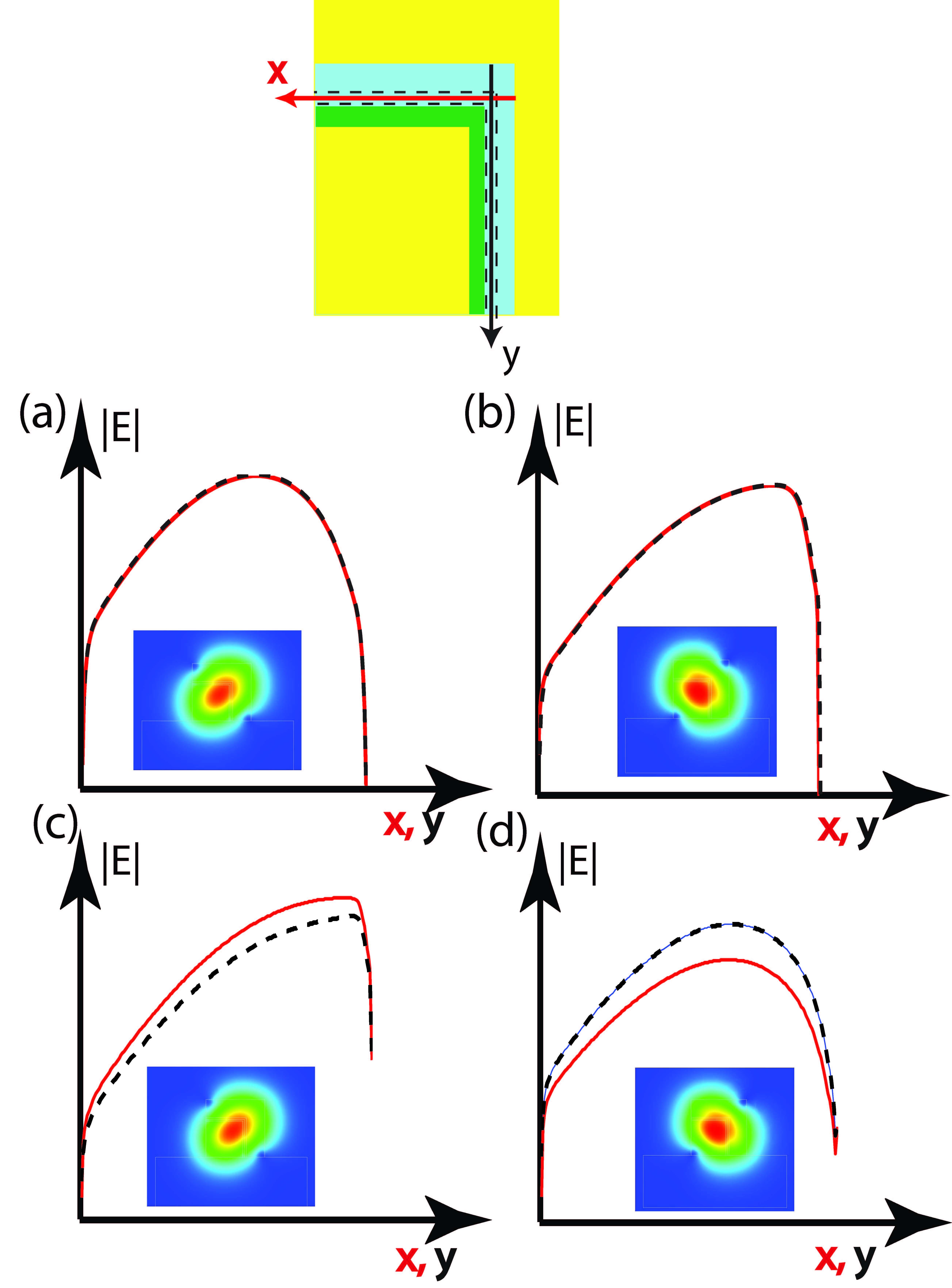}
\caption{\textbf{Electric field intensity in the accumulation layer at ENZ and non-ENZ biases}. \textbf{(a)}, \textbf{(b)} are the electric field intensities of the two rotated eigenmodes at $\mu=$-2.3 V. \textbf{(c)}, \textbf{(d)} are at $\mu=$-1.8 V. The red and black lines represent the intensity distributions in the horizontal and vertical accumulation layers respectively. \label{mode}}%
\end{figure}

The electric field intensity distribution within the accumulation layer is plotted in Fig. \ref{mode} to investigate the effect of ENZ on the eignemodes orientation angle. The red and black curves are the field intensity distributions in the accumulation region of the ITO layers in the horizontal and vertical directions, respectively. It is evident that the intensity distribution is diagonally symmetric, so the eignemodes would rotate 45$^{\circ}$ despite of the asymmetry of the structure (particularly the outer cladding) as seen in Figs. \ref{mode}(a) and (b). When biased away from the ENZ regime, the eigenmodes in the structure experience an asymmetric refractive index profile which is dictated by the structure. When the index of refraction of the low index region is tuned close to zero with the help of ENZ, the electric field intensity of the eigenmdoes are mostly concentrated within the slot region. This in turn reduces the asymmetry experienced by the eigenmodes of the structure, leading to a modulation in their rotation angle. In cases where we electrically gate the ITO outside the ENZ regime such as at a bias of $\mu=$-1.8 V in Fig. \ref{mode}(c) and (d), asymmetric field distribution forces the rotation angle to shift from 45$^{\circ}$. Accordingly, ENZ is key to achieving eigenmodes with an orientation of 45$^{\circ}$, which in turn enables the polarization insensitive attenuation. As highlighted in Fig. \ref{loss}(b), it is also worth noting that the eigenmodes are orthogonal to one another ($\theta_{1}+\theta_{2}=90^{\circ}$) so the energy splitting formula (\ref{orthogonal}) holds under all bias voltage values.
\subsection{Modulator performance and sensitivity analysis}
\begin{figure}[t!]
\centering
\includegraphics[width=0.49\textwidth]{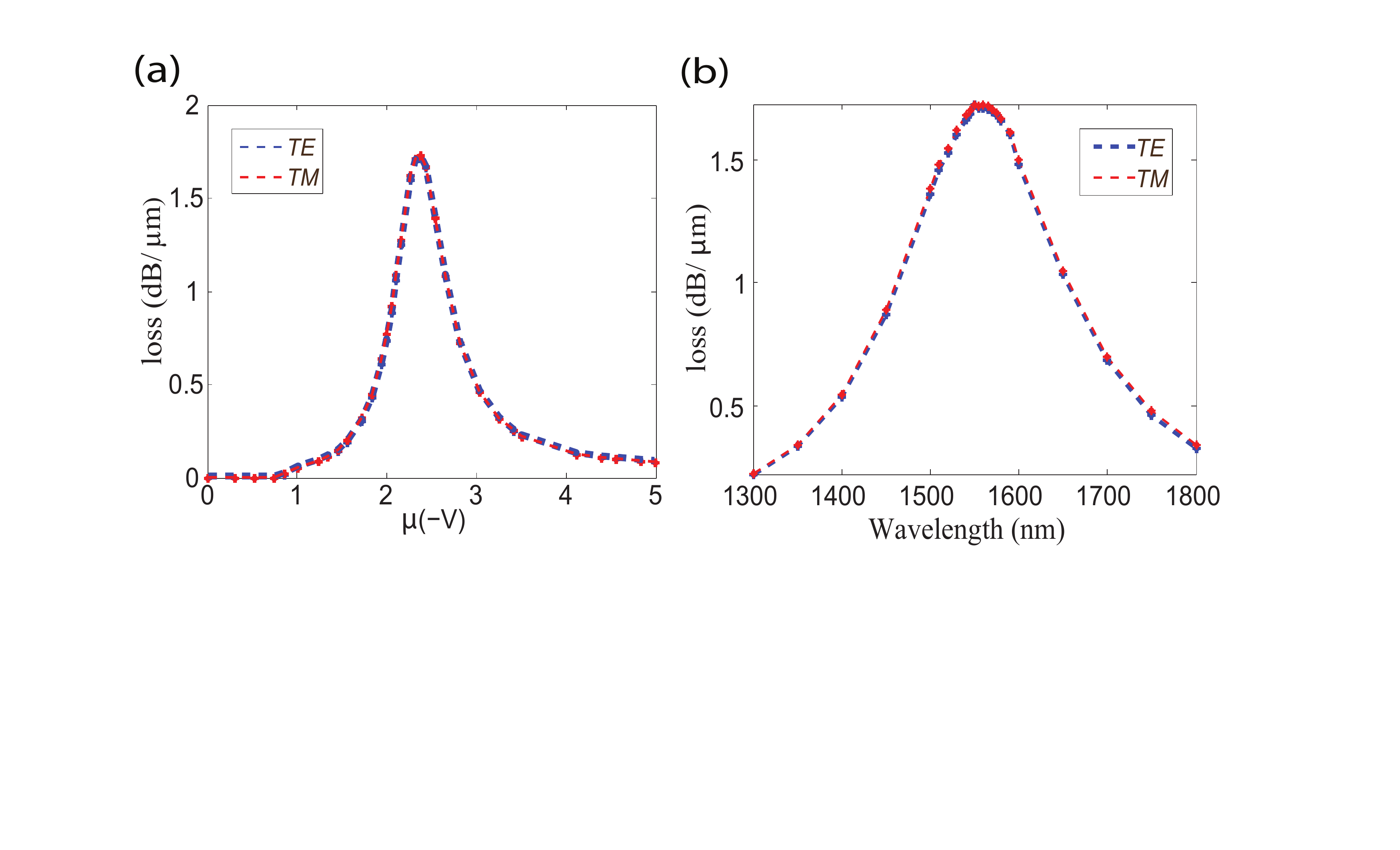}
\caption{\textbf{Attenuations of the TE and TM modes.} (a) Propagation losses of the TE and TM modes as function of gate voltage at operational wavelength of 1550 nm. (b) Wavelength dependance of attenuations of the TE and TM modes at $\mu=$-2.3 V. The full width at half maximum is 210 nm. \label{wavelength}}%
\end{figure}
To make a compact electroabsorption (EA) modulator, loss needs to be maximized for the OFF state to enhance the extinction ratio and minimized for the ON state to minimize the insertion loss. The extinction ratio (ER) is determined by:
\begin{equation}
{ER}\propto\frac{\min\left(\alpha_{1}^{OFF},\alpha_{2}^{OFF}\right)}{\max\left(\alpha_{1}^{ON},\alpha_{2}^{ON}\right)},
\end{equation}
where $\alpha$ refers to the propagation loss (in $dB/\mu$m) of the two eigenmodes for the ON and OFF states, respectively. Since the two eignemodes here serve as the basis for the input TE and TM modes, as indicated in (\ref{orthogonal}), the TE and TM propagation losses can therefore be expressed as:
\begin{equation}
\begin{array}{l}
\alpha_{TE} =10\log \left( {{{{10}^{0.1\alpha_{1}}}}{{\cos }^2}\theta  + {{10}^{0.1\alpha_{2}}}{{\sin }^2}\theta } \right)\\
\alpha_{TM} =10\log \left( {{{{10}^{0.1\alpha_{2}}}}{{\cos }^2}\theta  + {{10}^{0.1\alpha_{1}}}{{\sin }^2}\theta } \right),
\end{array}\label{mix}
\end{equation}
where $\alpha_{TE}$ and $\alpha_{TM}$ represent the losses of the input TE and TM polarizations, respectively. Figure \ref{wavelength}(a) plots the attenuation for both modes at different bias voltages. It shows that the two states of polarization virtually experience the same loss for all voltage values. This is achievable because of the 45$^{\circ}$ orientation of the eigenmodes that is achievable at $\mu=$-2.3 V, where ENZ takes place. For all other voltage values, where $\mu\ne$-2.3 V, the identical losses are also achievable because the losses for the two eigenmodes remain identical as seen in (\ref{mix}) and Fig. \ref{loss}. The attenuation reaches 1.71 $dB/\mu$m when biased within the ENZ regime and 0.052 $dB/\mu$m when $\mu=$-1 V, enabling a compact device footprint ($<$2 $\mu$m) for 3 $dB$ modulation depth. This results in a FOM as high as 31. The bias of such a structure involves providing a voltage difference between the ITO and the doped silicon core as shown schematically in Fig. \ref{schematic}(c).

\par For the modulator performance, the capacitance of the L-shaped slot layer of the $HfO_2$ buffer is calculated to be $C=$17.8 $fF$, and the corresponding voltage, which is required to switch from a maximum ($\mu=$-2.3 V) to a minimum attenuation ($\mu=$-1 V) is $\Delta V=1.3$ V. The resulting energy per bit is found to be $7.5$ $fJ$ with an electrical modulation bandwidth of 112 GHz. The operating bandwidth is determined by the RC time constant of the structure, where the resistance is estimated as $R=500 \Omega$ \cite{Sorger12}. This includes the resistance of $HfO_2$ and the intrinsic contact resistance \cite{Atwater10}. The wavelength dependence of the attenuation of the TE and TM modes is plotted in Fig. \ref{wavelength}(b), showing an optical bandwidth (full width at half maximum) of operation in excess of 210 nm (1450-1660 nm) with polarization insensitivity and a weak wavelength dependence.

\par
\begin{figure}
\centering
\includegraphics[width=0.36\textwidth]{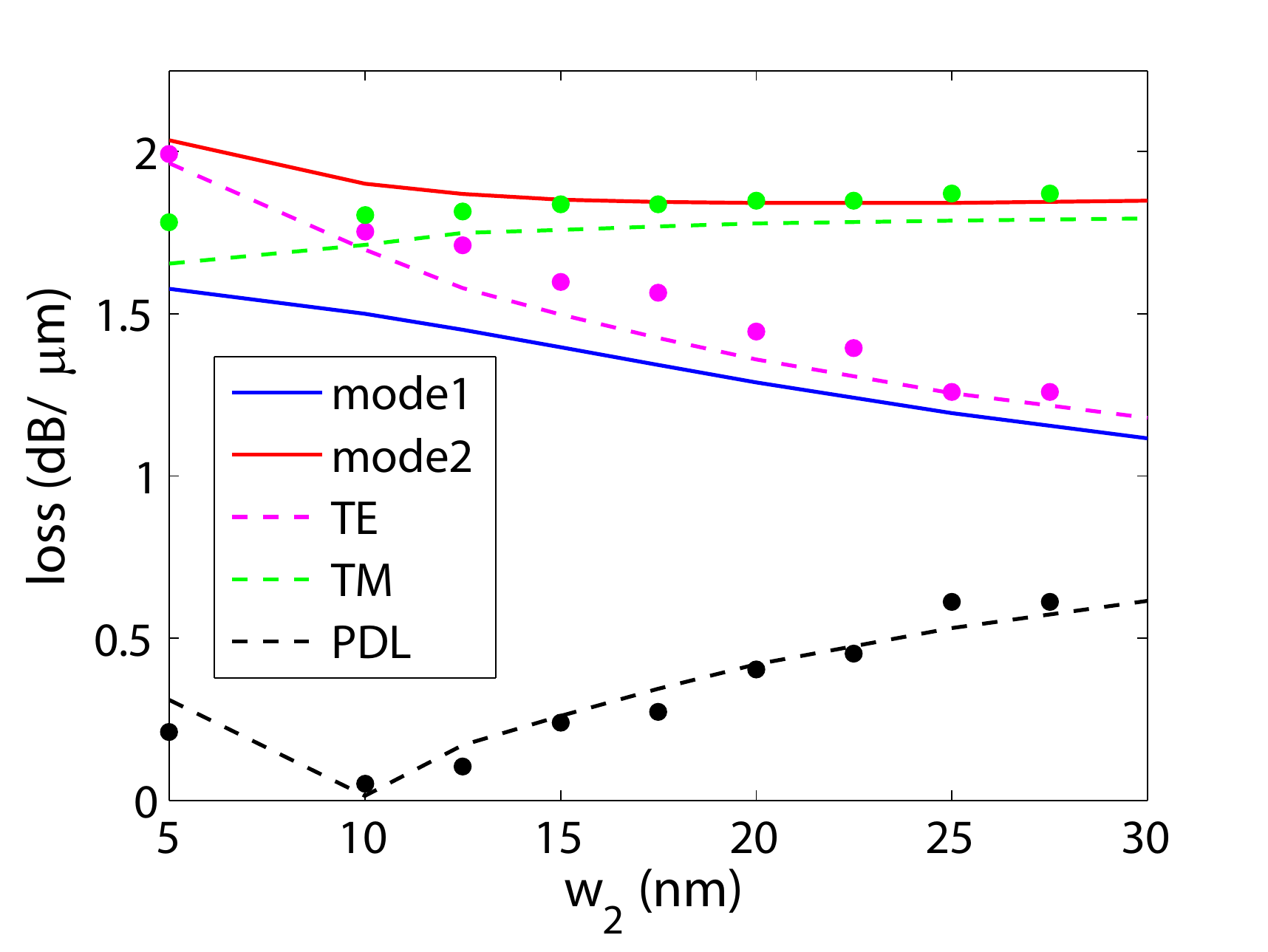}
\caption{\textbf{Mode attenuations and PDL under fabrication error.} Waveguide losses as $w_{2}$ ranges from 5 nm to 30 nm. The solid curves are the losses of the two rotated eigenmodes. The dashed purple and green curves are calculated by (\ref{mix}). The purple and green points are the simulated FDTD propagation losses under TE and TM light incidences respectively.\label{sweep}}%
\end{figure}

 Fabrication will inevitably introduce some errors thus breaking the proposed quasi-symmetric waveguide configuration. Figure \ref{sweep} plots the attenuation of the two eigenmodes and the TE-TM polarization dependent loss (PDL) as the width $w_2$ varies from 5 nm to 30 nm. The dashed purple and dashed green curves are attenuation values for the TE and TM modes obtained according to (\ref{mix}). It is evident that the polarization insensitive condition occurs as $w_{2}$ around 10 nm, which is the condition for the orientation of the eigenmodes to be equal to 45$^{\circ}$. For larger $w_{2}$, the diagonal quasi-symmetry is severely broken such that the eigenmodes become quasi-TE and quasi-TM modes, with $\theta=$16$^{\circ}$ at $w_{2}=$30 nm. To estimate the acceptable fabrication tolerance, the error margin is defined as:
\begin{equation}
PDL<0.16\min\left({{\alpha}_{TE}},{{\alpha}_{TM}}\right),\label{margin}
\end{equation}
where $0.16$ is obtained by $10\log(0.9)/3$, which translates to the condition that the polarization transmission power difference is less than 0.46 $dB$ (90\%) for 3 $dB$ modulator. The error margin in our design allows for 15 nm tolerance in the layer thickness, which is well within capabilities of existing nano-fabrication tools. These results are supported by FDTD simulations. For the present case the polarization independent modulator is achieved for $w_{2}=$9 nm, which is close to the theoretical prediction of 10 nm. PDL also agrees well with the theoretical calculations and allows for 15 nm tolerance. It should be highlighted that the design methodology we describe here does not require any symmetry in the structure. The key to the performance is the addition of a tunable material in a waveguide system, where the eigenmodes are rotated by 45$^{\circ}$. This point is demonstrated in the waveguide geometry shown in Fig. \ref{schematic}(b), where the two eigenmodes are oriented at an angle of 45$^{\circ}$ and exhibit propagation loss of 0.038 and 0.035 $dB$/$\mu$m respectively as $\mu=$-5 V ($\varepsilon$=-4.4-1.23$i$ \cite{Vasudev13}). Calculations using the EME approach predict that the TE and TM inputs suffer from losses of 0.036 and 0.0365 $dB/\mu$m, which are values that reside between the losses for the two eigenmodes. This can be expected due to the field decomposition of the TE and TM modes into the waveguide eigenmodes.

\par When a polarized beam is input to the L-slot polarization rotator, the coupling efficiency can be optimized by matching the spatial and phase profiles of the modes of the input waveguide to the rotated eigenmodes of the rotator \cite{Morichetti07}. The cross section of the input waveguide is therefore preferred to be engineered similar to that of the L-slot modulator for the maximal spatial overlap of the field. For the phase matching condition, the effective index of the input mode supported by the input waveguide is desired to fall between the indices of the two rotated eigenmodes of the L-slot modulator so they can be equally excited. In addition, the rotator-based modulator may induce a change in the state of polarization at the output of the modulator depending on the length of the modulator being used. The degree of the change in the output polarization state can be estimated by using the polarization conversion efficiency (PCE) of a polarization rotator \cite{Deng05}:
\begin{equation}
PCE={{\sin }^2}(2\theta){{\sin}^2}(\frac{{\pi}{L_{m}}}{2L_{\pi}}),
\end{equation}
where $L_m$ is the length of the modulator. For a modulator in which ITO is biased within ENZ regime, the length of the modulator ($<$2 $\mu$m) is much less than the coupling length (38 $\mu$m) so PCE is very close to 0, meaning that the polarization state at the output of the modulator is almost identical to that at the input. However, if the length of the modulator is comparable to the coupling length, the input state of the polarization would undergo a measurable rotation as the beam propagates through the modulator. Therefore, the device structure can be engineered such that the length of the modulator is equal to even integer of $L_{\pi}$ or cascaded by a passive rotator so that the output state of polarization can be maintained.
\subsection{Nanoscale polarization insensitive modulator}
The extinction ratio of the rotator modulator can be increased via the enhancement of the electric field within the absorbing medium. In the case of the slot waveguide architecture, this is usually achieved by either decreasing the slot layer thickness or increasing the refractive index contrast between the ITO and the adjacent materials. Assuming the thickness of $HfO_2$ and ITO to be 2 nm respectively, the reduction in the thickness of the buffer layer leads to the increase in the absorption loss of the two eigenmodes to be 2.6 and 2.39 $dB/\mu $m. In addition, if one increases the index contrast of the modulator waveguide such as replacing the $HfO_2$ by $TiO_2$ (n=2.37 \cite{Bradley12}), the losses can be further raised to 2.71 and 2.57 $dB/\mu $m. As such the device can reach a nanoscale footprint (1130 nm in length) while at the same time it can maintain its polarization independent property when used as the basis for a modulator. In this case the modulator would have an extinction of 2.64 $dB/\mu$m. For this device where the resistance is assumed to be $R=500 \Omega$ \cite{Sorger12}, the capacitance is calculated to be 66.3 $fF$ ($\varepsilon_r$=31 for $TiO_2$ \cite{Tang94}), which results in energy per bit of 28 $fJ$ and electrical modulation bandwidth of 30.2 GHz.
\section{Conclusion}
In conclusion, we report a systematic and generic design methodology to achieve various polarization engineering functions using lossless media. Through the utilization of an L-slot architecture the polarization state rotator demonstrated here can be realized using nanoscale cross section and a length of only a few microns, with no use of metal layers. These device sizes are at least 4 times smaller than existing designs, which helps address the rising demand of higher packing density in photonic integrated circuit. If this L-slot design is utilized in a hybrid coupler with another silicon nanowire, the rotator coupler then enables versatile polarization management functions, where light with arbitrary state of linear polarization can be separated, generated or eliminated. In addition, incorporating a tunable material into the structures discussed here enables the realization of energy efficient polarization insensitive modulators. This can be achieved using a simple and systematic design process with no need for numerical geometry sweeping approaches. Such modulators can be readily realized using ITO layers added to the L-slot region, which enables one to utilize the ENZ effects exhibited by ITO. To demonstrate the universality of this methodology, an example has been discussed, where there is no symmetry in the waveguide structure, however the waveguide is tuned to ensure that the eigenmodes are rotated by 45$^{\circ}$. In this example, we show how a polarization insensitive modulator can be realized with a modulation bandwidth of 112 GHz, a length of 1800 nm, an energy per bit of 7.5 $fJ$ and an optical bandwidth of 210 nm. Furthermore, by utilizing a tunable material such as ITO and the hybrid coupler architecture, devices for dynamic polarization control can be realized. More importantly even though the rotator designs discussed here are illustrated in SOI, the overall design methodology features strong flexibility and is easily implemented in other material platforms.

\ifCLASSOPTIONcaptionsoff
 \newpage
\fi


\begin{IEEEbiographynophoto}{PoHan Chang}
received the B.A.Sc. degree in 2010 and the M.A.Sc. degree in 2012 from National Taiwan University. He is now working toward Ph.D. degree in Department of Electrical and Computer Engineering at University of Toronto, Canada. His research interests include plasmonics waveguide and its applications, metamaterial-assisted waveguide structure, and electromagnetic wave propagation.
\end{IEEEbiographynophoto}

\begin{IEEEbiographynophoto}{Charles Lin}
received the B.A.Sc. degree in 2010 and the M.A.Sc. degree in 2012, both in Electrical Engineering from the University of Toronto, where he is currently working toward the Ph.D. degree in photonics. His research interests include plasmonic resonance and field effects, plasmonic waveguiding structures, and nanoplasmonic devices and applications.
\end{IEEEbiographynophoto}


\begin{IEEEbiographynophoto}{Amr S. Helmy}
is a Professor in the department of electrical and computer engineering at the University of Toronto. Prior to his academic career, he held a position at Agilent Technologies photonic devices, R\&D division, in the UK between 2000 and 2004.  He received his Ph.D. and M.Sc. from the University of Glasgow with a focus on photonic devices and fabrication technologies, in 1999 and 1995 respectively.  He received his B.Sc. from Cairo University in 1993, in electronics and telecommunications engineering.

He has published in IEEE and OSA Journals and Conferences. His research interests include photonic device physics and characterization techniques, with emphasis on nonlinear optics in III-V semiconductors; applied optical spectroscopy in III-V optoelectronic devices and materials; III-V fabrication and monolithic integration techniques.

Amr is a senior member of the IEEE and the Optical Society of America. He is currently the program chair of the IEEE Photonics Conference, the chair of the semiconductor laser subcommittee for the conference on lasers and electro-optics (CLEO) as well as an associate editor of the IEEE Photonics Journal and Optics Express. Amr has also served as VP membership for the IEEE LEOS between 2007 and 2010.
\end{IEEEbiographynophoto}
\end{document}